\documentclass[prl,aps,twocolumn,showpacs,floatfix]{revtex4}

\usepackage{mathrsfs}
\usepackage{graphicx}           
\usepackage{amsmath,amssymb}
\usepackage{version}

\DeclareGraphicsExtensions{.eps,.ps}

\newcommand{\bQ}{{\bf Q}}
\newcommand{\bk}{{\bf k}}

\newcommand{\br}{{\bf r}}

\newcommand{\beqa}{\begin{eqnarray}}
\newcommand{\eeqa}{\end{eqnarray}}

\begin{document}

\title
{Dielectric response effects in attosecond time-resolved streaked
photoelectron spectra of metal surfaces}
\author{C.-H. Zhang and U. Thumm}
\affiliation{Department of Physics, Kansas State University,
Manhattan, Kansas 66506, USA}
\date{\today}

\begin{abstract}
The release of conduction-band electrons from a metal surface by a
sub-femtosecond extreme ultraviolet (XUV) pulse, and their
propagation through the solid, provokes a dielectric response in the
solid that acts back on the photoelectron wave packet.  We
calculated the (wake) potential associated with this photoelectron
self-interaction in terms of bulk and surface plasmon excitations
and show that it induces a considerable, XUV-frequency-dependent
temporal shift in laser-streaked XUV photoemission spectra,
suggesting the observation of the ultrafast solid-state dielectric
response in contemporary streaked photoemission experiments.
\end{abstract}


\maketitle

The sudden release and subsequent motion of a photoelectron (PE) in
and in front of a solid dielectric medium provokes collective
electron excitations in the solid. The back-interaction of these
excitations with the PE can be modeled as a dynamic, wave-like
redistribution of electronic density in the solid in terms of a
complex-valued effective electron-self-interaction (or ``wake")
potential~\cite{Harris73,Manson81,Echenique90,Garcia92}.
This density wake appears since the electron distribution in the
solid cannot equilibrate during the motion of the released electron.
The wake potential depends on the kinetic energy $E$ of the PE. Its
real part is due to virtual excitations of bulk and surface
plasmons, while its imaginary part accounts for inelastic scattering
and energy loss. The dependence of the dynamic wake potential on the
charge state and velocity of a classical particle has been studied
extensively over several decades with regard to energy
loss~\cite{Lucas70,Sunjic71}, electron-exchange and -correlation
contributions~\cite{Overhauser71}, electron emission in ion-surface
collisions~\cite{Burgdorfer}, and electronic self-interaction
effects on photoelectron spectra~\cite{GersTzoar73}. While these
examples emphasize the influence of the solid's dielectric response,
they do not resolve the ultrafast electronic response in the
condensed-matter-plus-charged-particle system in time. Owing to
significant progress in laser technology over the past decade,
sub-femtosecond XUV pulses can now be generated and synchronized
with the primary IR laser pulse, allowing the time-resolved
observation of the electronic dynamics in atoms~\cite{Schultze10}
and solids~\cite{Cavalieri07}. Time-resolved experiments at the
intrinsic time-scales of an active electron and the correlated
dynamics of two electrons~\cite{Drescher02,Miaja08} or
plasmons~\cite{Kubo05,Stockman07} promise unprecedented sensitive
experimental tests of collective electronic transport phenomena in
solids and novel plasmonic devices~\cite{Sukharev06,Nordlander07}.

Using attosecond time-resolved XUV+IR pump-probe technology in a
proof-of-principle experiment, a relative temporal shift of 110 as
(1 as =$10^{-18}$ s) between the photoemission of 4f core level and
conduction-band electrons from a tungsten (110) surface was
measured~\cite{Cavalieri07}. Essential for the correct reproduction
of this shift within simple quantum mechanical
models~\cite{Zhang09,Zhang10} is the proper inclusion of (i) the
PE's phase evolution during the streaked emission, (ii) the
attenuation of the IR pulse inside the solid (skin effect), and
(iii) electron transport effects in the solid. The phase of the PE
is further affected by its plasmon-mediated self-interaction while
moving inside and outside the solid, and we expect streaked
photoemission experiments to help reveal these dynamic many-body
effects in solids. In this Letter we investigate how these three
factors affect the PE dynamics and extend previous theoretical
studies on the streaked photoemission from solid
surfaces~\cite{Baggesen09,Lemell09,Kazansky09,Zhang09} to expose the
effect of the dynamic plasmon response on time and energy
resolved PE spectra.

We calculate the dynamic wake potential assuming that the released
PE moves with a constant velocity $v_z>0$ along a classical
trajectory towards and perpendicular to the surface and crosses the
metal-vacuum interface ($z=0$) at time $t = 0$, leading to the
density $\rho(\br)=\delta(\br_{\parallel})\delta(z-v_zt)$
(throughout this Letter we use atomic units, unless stated
otherwise). The semi-infinite solid is modeled in jellium
approximation~\cite{Zhang09}, and its excitations are described by
the dispersion relations for bulk- and surface-plasmon excitations,
$\omega^2_{k}=\omega_p^2+3k_F^2k^2/5+k^4/4$ and
$\omega^2_{Q}=\omega_s^2+\sqrt{3}k_F\omega_sQ/\sqrt{5}+\beta
Q^2+Q^4/4$~\cite{Echenique81,Garcia92}. For low momenta $\bk=(\bQ,
k_z)$, these relations model single-plasmon modes with bulk- and
surface-plasmon frequencies $\omega^2_p=4\pi n$ and $\omega_s
=\omega_p/\sqrt{2}$, respectively,
 and decay into particle-hole
excitation at high momenta through the terms $k^4/4$ and $Q^4/4$.
$n$ is the bulk-conduction-electron density, and
$k_F=(3\pi^2n)^{1/3}$ is the Fermi velocity. $\beta$ is determined
so that the surface-plasmon dispersion relation joins the
particle-hole continuum at the same point as the bulk
line~\cite{Echenique81}. The plasmon field of the solid is then
given by the Hamiltonian
$H_0=\sum_{\bQ,k_z\ge0}\omega_kb^\dagger_{\bk}b_{\bk}
+\sum_{\bQ}\omega_{Q}a^\dagger_{\bQ}a_{\bQ}$, where
$b^{(\dagger)}_{\bQ}$ and $a^{(\dagger)}_{\bk}$ are annihilation
(creation) operators for bulk and surface plasmons,
respectively~\cite{Lucas70,Overhauser71,GersTzoar73}. The
interaction between the PE and the jellium solid is given by
$H_{1}=\int d\br \rho(\br)
\left[\phi_{b}(\br)+\phi_{s}(\br)\right]$,
where $\phi_b(\br)=\sum_{\bQ,k_z\ge0}B_{\bk}b_{\bk}\sin(k_zz)
e^{i\bQ\cdot\br_{\parallel}}\Theta(-z)+\mbox{h.c.}$ and
$\phi_s(\br)=\sum_{\bQ}A_{\bQ}a_{\bQ}e^{-Q|z|}
e^{i\bQ\cdot\br_{\parallel}}+\mbox{h.c.}$ are the bulk and surface
plasmon fields. $|B_{\bk}|^2= 8\pi\omega_p^2 / (V k^2\omega_k)$ and
$|A_{\bQ}|^2= \pi\omega_s^2 / (S Q\omega_Q)$ are the interaction
strengths (with quantization volume $V$ and surface
$S$)~\cite{Overhauser71}.

The wavefunction $|\psi(t)\rangle$ of the bulk and surface plasmon
field interacting with the classical charged particle for the model
Hamiltonian $H=H_0+H_1$  can be solved
exactly~\cite{Lucas70,Sunjic71}. The wake potential of the PE is
then calculated as
$V_{im}(t)=\frac12\langle\psi(t)|H_{1,I}(t)|\psi(t)\rangle$, where
$H_{1,I}(t)$ is the interaction-picture presentation of $H_1$.
Following \cite{Overhauser71,Echenique81,Garcia92},
we obtain the real part of wake potential
\begin{widetext}
\begin{align}
\label{eq:imp} V_{im}^{r}(z,v_z)=&
\frac{\Theta(-z)\omega_p^2}{\pi}\int_0^{\infty}\!\!\frac{dk}{k}
\int_0^k\!\!dk_z\frac{(k_z^2v_z^2-\omega_k^2)[1-\cos(2k_zz)]}
{(k^2_zv^2_z-\omega_k^2)^2+k_z^2v_z^2\gamma^2}
\displaybreak[0]\nonumber\\& -\Theta(z)v_z\omega_s^2
\int_0^{\infty}\!\!dQQ\frac{e^{-Qz}}{\omega_Q}
\frac{\sin(\omega_Qz/v_z)}{Q^2v_z^2+\omega^2_Q}
-\frac{\omega_s^2}{2}\int_0^{\infty}\!\!dQ\frac{e^{-2Q|z|}}{Q^2v_z^2+\omega^2_Q},
\end{align}
\end{widetext}
where $\Theta(z)$ is the unit step function and $\gamma$ the decay-width
of the plasmon excitation. The first term in (\ref{eq:imp}) includes
bulk and the last two terms surface plasmon excitations.

\begin{figure}[t]
\begin{center}
\includegraphics[width=1.0\columnwidth,keepaspectratio=true,
draft=false]{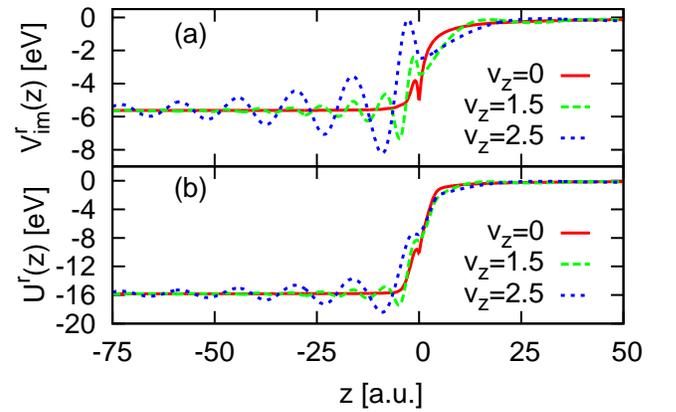}  \vspace{-6mm} \caption{(Color online)
(a)Dynamic wake potential $V_{im}^{r}(z,v_z)$ for Al for different
PE velocities $v_z$. (b) Real part of $U(z,v_z)$, adjusted to the
Fermi energy of Al at $v_z=0$. \label{fig:image_pot} } \vspace{-6mm}
\end{center}
\end{figure}

Figure~\ref{fig:image_pot}(a) shows $V_{im}^{r}(z,v_z)$ for
different PE velocities $v_z$ for aluminum with
$\omega_s=0.378$~\cite{Ibach06}, assuming $\gamma=0.1\omega_s$. The
static image potential is obtained in the limit $v_z=0$. For $v_z >0
$, $V_{im}$ oscillates near the metal surface with wavelength
$\lambda_b=\pi v_z/\omega_{p}$ inside the solid and
$\lambda_{s}=2\pi v_z/\omega_s$ in the vacuum, and approaches
$-1/4z$~\cite{Ibach06} far away from the surface for all $v_z$, as
expected. Equation (\ref{eq:imp}) underestimates the influence of
the positively charged ion cores, and its bulk limit ($\approx
5.8$~eV) does not reproduce the Al Fermi energy ($\varepsilon_F
=11.7$~eV)~\cite{Ashcroft76}. We therefore add a step potential to
obtain the effective dynamic image potential,
\begin{align}
U(z,v_z)=-\frac{V_0}{1+e^{z/a}}+  V_{im}(z,v_z)
\end{align}
and adjust the depth to $V_0=11.7$~eV and interface-thickness
parameter to $a=1.4$~\AA, respectively, in order to reproduce
$\varepsilon_F$ in the static limit for $v_z=0$
[Fig.~\ref{fig:image_pot}(b)].

\begin{figure}[t]
\begin{center}
\includegraphics[width=1.0\columnwidth,keepaspectratio=true,
draft=false]{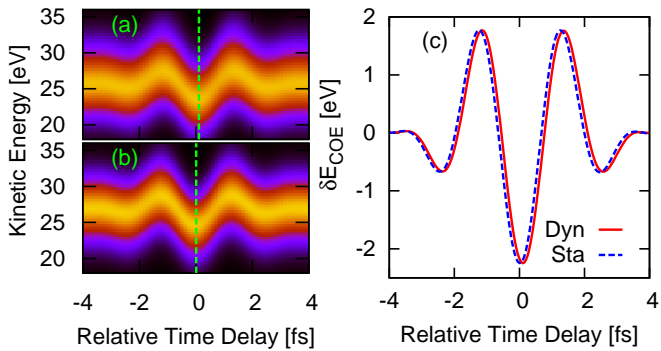}  \vspace{-6mm} \caption{(Color online)
Streaked PE spectra for an Al surface calculated with the (a)
dynamic image potential $\delta U(z,v_z)$, (b) static image
potential $\delta U(z,v_z=0)$ for $\hbar\omega_X=40$~eV, skin depth
$\delta_L=0$, and mean free path $\lambda=5$~\AA. (c) Corresponding
center-of-energy shifts $\delta E_{COE}$.  The temporal shift
between the  traces $\delta E_{COE}$ for dynamic and static image
potentials is $\approx 100$~as. \label{fig:exuv40_spect_coe} }
\vspace{-6mm}
\end{center}
\end{figure}

We model the metal surface as a $300$~a.u wide slab and obtain its
eigenvectors $\varepsilon_n$ and and wave functions $\psi_n(z)$ by
diagonalizing the time-independent Schr\"odinger equation (SE)
\begin{align}
\label{eq:gs}
\varepsilon_n\psi_n(z)&=\left[-\frac{1}{2}\frac{d^2}{dz^2}+U(z,v_z=0)\right]\psi_n(z).
\end{align}
In typical streaking experiments, the XUV pulse intensity is
sufficiently low, so that photoemission in the XUV pulse $E_X(t)$
can be treated perturbatively. The release and propagation of the PE
wavepacket $\delta\psi_n(z,t)$ from the state $\psi_n(z,t)$ is then
dictated by the SE~\cite{Zhang10}
\begin{align}
\label{eq:ex} i\frac{\partial}{\partial
t}\delta\psi_n(z,t;\tau)=&\left[H_L(t)+\delta U(z,v_z)\right]
\delta\psi_n(z,t;\tau)
\displaybreak[0]\nonumber\\
&+zE_X(t+\tau)\psi_n(z,t;\tau).
\end{align}
$H_L=\frac12\left[-i\frac{d}{dz}+A_L(z,t)\right]^2+U(z,0)$ is the
Hamiltonian for the solid slab in the presence of the IR-laser pulse
$A_L(z,t)$, $\tau$ the delay between the XUV and IR pulses with
$\tau>0$ if the XUV precedes the IR pulse, $\delta U(z,v_z) =
V_{im}(z,v_z)-V_{im}(z,0)$ represents the complex-valued dynamic
part of the PE self-interaction, and
$v_z=\sqrt{2(\omega_X-|\varepsilon_n|)}$ is the PE speed. The
evolution of the $n$-th initial state of the slab below the Fermi surface in the
IR pulse is given by
\begin{align}
\label{eq:init} i\frac{\partial}{\partial t}\psi_n(z,t)
=H_L(t)\psi_n(z,t).
\end{align}
Since $E_{L(X)}(t\rightarrow\pm\infty)=0$, (\ref{eq:ex}) and
(\ref{eq:init}) are solved for the initial conditions
$\delta\psi_n(z,t\rightarrow-\infty;\tau)=0$ and
$\psi_n(z,t\rightarrow-\infty)=\psi_n(z)e^{-i\varepsilon_nt}$.

We represent the vector potential of the laser pulse as
$A_{L}(t)=A_{0}\sin^2\left(\pi
t/\tau_L\right)\cos\left[\omega_{L}\left(t-\tau_L/2\right)\right]$
for $0\le t\le \tau_L$ and 0 otherwise, with
$\hbar\omega_{L}=1.57$~eV, intensity $I_{L}= A_0^2\omega_L^2/2=
5\times10^{11}$W/cm$^2$, and pulse length $\tau_{L}=8$~fs. We assume
an exponential damping of the IR laser field inside the solid
$A_L(z,t)=A_L(t)\left[e^{z/\delta_L}\Theta(-z)+\Theta(z)\right]$
with a skin depth $\delta_L$, and take a Gaussian XUV pulse with
pulse length $\tau_{X}=300$~as.

Assuming free-electron dispersion, $E=k^2/2$, the
energy-differential photoemission probability
$P(E,\tau)=\sum_{\varepsilon_n<\varepsilon_F}
\left|\delta\tilde{\psi}_n(k,\tau)\right|^2$
leads to the delay-dependent center of energy (COE) of the PE
spectrum
\begin{align}
\label{eq:Ecom}
E_{COE}(\tau)=\frac{1}{2P_{tot}(\tau)}\sum_{\varepsilon_n<\varepsilon_F}\!\!\int
\!\!dk\left|k \, \delta\tilde{\psi}_n(k,\tau)\right|^2,
\end{align}
where $\delta\tilde{\psi}_n(k,\tau)$ is the Fourier transform of
$\delta\psi_n(z,t\rightarrow\infty;\tau)$ and
$P_{tot}(\tau)=\sum_{\varepsilon_n<\varepsilon_F}\int dk
\left|\delta\tilde{\psi}_n(k,\tau)\right|^2$ the total emission
probability.

\begin{figure}[t]
\begin{center}
\includegraphics[width=1.0\columnwidth,keepaspectratio=true,
draft=false]{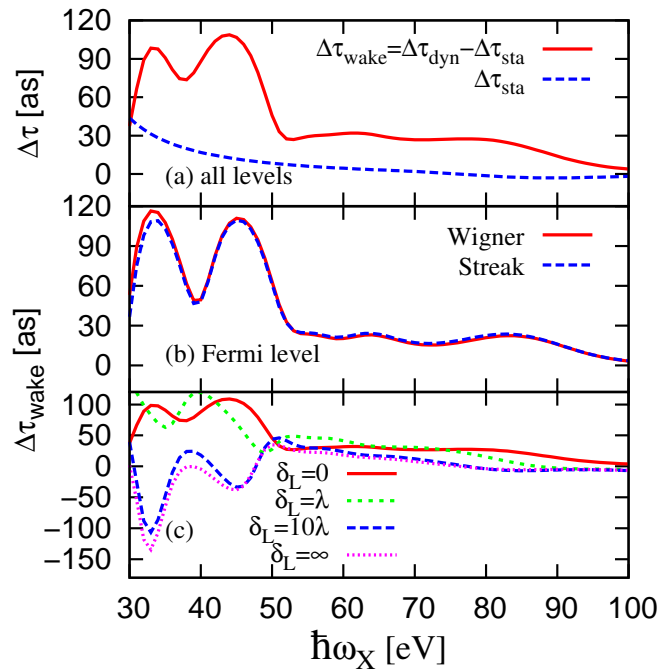}  \vspace{-6mm} \caption{(Color online) (a)
$\Delta\tau_{wake}$ and $\Delta\tau_{sta}$ as a function of the XUV
frequency obtained by fitting Eq.~(\ref{eq:Ecom_fit} ) to the
calculated $E_{COE}(\tau)$, including all levels below the Fermi
surface, for skin depth $\delta_L=0$. (b)
Comparison of $\Delta\tau_{wake}$, obtained from the streaked
$E_{COE}(\tau)$ at the Fermi surface, and the Wigner delay
$\Delta\tau^W_{wake}$. (c)$\Delta\tau_{wake}$ for
different skin depths. \label{fig:fig3} } \vspace{-6mm}
\end{center}
\end{figure}

In order to reveal the effect of the dynamic dielectric response on
the streaked PE spectrum, we compare the results of two separate
calculations for values of $\hbar\omega_X$ between 30 and 100~eV.
First, we solve (\ref{eq:ex}) without $\delta U(z,v_z)$ and denote
the results as ``static". Next, we add wake effects and obtain
``dynamic" results by including the real part of $\delta U(z,v_z)$
in (\ref{eq:ex}). In both calculations, we replace the imaginary
part of $\delta U(z,v_z)$ by the phenomenological expression $\delta
U^i_{ph}(z,v_z)=-v_z\Theta(-z)/ (2\lambda)$~\cite{notez}, where
$\lambda$ is the PE mean-free path.

We first present results for $\delta_L=0$ for which the IR field is
completely screened inside the solid. Figure
\ref{fig:exuv40_spect_coe} compares the static and dynamic streaked
PE spectra and their center of energies for $\hbar\omega_X=40$~eV.
We define temporal shifts $\tau_{sta(dyn)}$ for static (dynamic)
calculations relative to $A_L$ by fitting~\cite{Zhang09,Zhang10}
\begin{align}
\label{eq:Ecom_fit}
E^{sta(dyn)}_{COE}(\tau)=a+bA_L(\tau-\Delta\tau_{sta(dyn)}).
\end{align}
Fig.~\ref{fig:exuv40_spect_coe}(c) shows the relative temporal shift
$\Delta\tau_{wake}=\Delta\tau_{dyn}-\Delta\tau_{sta}\approx100$~as,
suggesting a noticeable - on the scale of the temporal resolution in
measured streaked PE spectra - contribution of the dynamic plasmon
response to the temporal shift.

Our results for $\Delta\tau_{dyn}$ and $\Delta\tau_{sta}$ as a
function of $\hbar\omega_X$  in Fig.~\ref{fig:fig3}(a) reveal that
the dynamic wake potential has a significant effect on the PE delay,
especially for $\hbar\omega_X<$50~eV, where $\Delta\tau_{dyn}$
develops a double-hump structure. The $\hbar\omega_X$ dependence of
$\Delta\tau_{wake}$ can be understood as due to scattering of the PE
in the wake potential $\delta U(z,v_z)$. This interaction of the PE
with $\delta U(z,v_z)$ changes the phase of $\delta\psi_n(z,t)$,
giving rise to a Wigner delay $\Delta\tau^W$~\cite{Carvalho02}. We
determine $\Delta\tau^W_{sta(dyn)}$ for $A_L=0$ by relating the PE
position $\langle z\rangle=\int dkz|\delta\psi_{n}(z,t)|^2$ and
velocity $\langle v\rangle=\int dk|k\delta\psi_{n}(k,t)|^2$ at a
time $t \gg \tau_L$ according to $\langle z\rangle=\langle v\rangle
(t-\Delta\tau^W)$. In support of this ``scattering interpretation",
Fig.~\ref{fig:fig3}(b) shows excellent agreement of the streaking
delay $\Delta\tau_{wake}$ with
$\Delta\tau^{W}_{wake}=\Delta\tau^{W}_{dyn}-\Delta\tau^{W}_{sta}$
for photoemission from the Fermi level. We find equally good
agreement for emission from initial conduction band states below the
Fermi level. For this comparison, we assumed that the IR field does
not penetrate the solid ($\delta_L=0$).

Since the effect of the actual IR skin depth on the streaked
spectrum from surfaces is currently debated
~\cite{Cavalieri07,Kazansky09,Zhang09,Lemell09,Baggesen09}, we found
it compelling to study the influence of $\delta_L$ on
$\Delta\tau_{wake}$. Our numerical results in Fig.~\ref{fig:fig3}(c)
show that $\Delta\tau_{wake}$ is strongly affected by changes in
the IR skin depth for $\delta_L \lesssim 2\lambda$, due to AC Stark
polarization of the initial states $\psi_{n}(z,t)$ and the PE wave
packet $\delta\psi_n(z,t)$, as well as the combined action of
$U(z,v_z)$ and $A_L(z,t)$ on $\delta\psi_n(z,t)$~\cite{Zhang10}. As
contributions to the emitted photocurrent are limited to PEs that
are released within a few mean-free paths $\lambda$ from the
surface, $\Delta\tau_{wake}$ becomes less IR skin-depth dependent
for $\delta_L \gtrsim 2\lambda$ and stabilizes in the limit
$\delta_L \rightarrow \infty$.

\begin{figure}[t]
\begin{center}
\includegraphics[width=1.0\columnwidth,keepaspectratio=true,
draft=false]{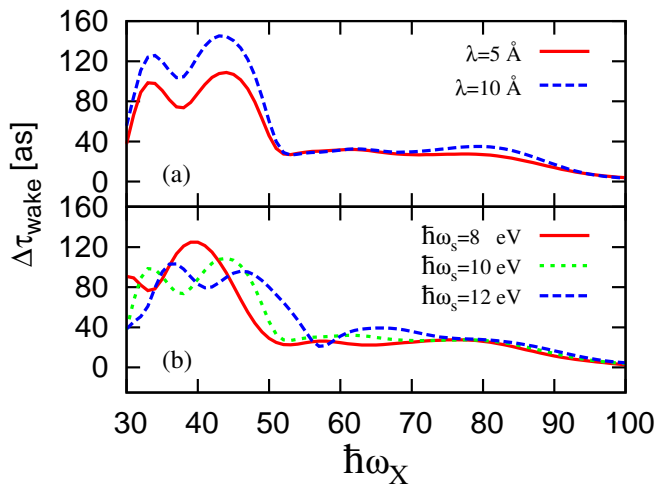}  \vspace{-6mm} \caption{(Color online)
Relative delay $\Delta\tau_{wake}$ as a function of the XUV
frequency at different (a) PE mean-free paths and (b) surface
plasmon frequencies.\label{fig:fig4}} \vspace{-6mm}
\end{center}
\end{figure}

Figure~\ref{fig:fig4} shows the dependence of $\Delta\tau_{wake}$ on
the PE mean free path and the surface plasmon frequency for
$\delta_L=0$. Increasing $\lambda$ by a factor of two significantly
increases $\Delta\tau_{wake}$ for $\hbar\omega_X \lesssim 50$~eV,
but has little influence at larger $\hbar\omega_X$
(Fig.~\ref{fig:fig4}(a)). Our result that, in general,
$\Delta\tau_{wake}(2\lambda)\ne\Delta\tau_{wake}(\lambda)$, is
incompatible with the interpretation~\cite{Cavalieri07,Kazansky09}
of the delay between photoemission from core and conduction-band
levels in tungsten being due solely to the PE's average travel time
in the solid ($\approx \lambda/\langle v\rangle$). Decreasing
$\omega_s$ shifts the double-hump structure to the lower
$\hbar\omega_X$, and thus to lower kinetic energies of the PEs, as
expected in view of the decreased thresholds for surface and bulk
plasmon excitation (Fig.~\ref{fig:fig4}(b)).

In conclusion, we have calculated the wake potential induced in an
aluminum surface by a PE that is released in the electric field of an
attosecond XUV pulse. By comparing centers of energies and
photoemission delays in IR-streaked PE spectra including the dynamic
wake potential with calculations performed in the static limit, we
find a significant contribution to the temporal shift
$\Delta\tau_{wake}$ in photoemission from the metal conduction band.
This shift is due to the excitation of the bulk and surface plasmons
in the metal during photoemission and is found to sensitively depend
on the XUV frequency as well as on solid state characteristics, such
as the bulk (surface) plasmon frequency, IR skin depth, and
PE transport in the solid. The measurement of streaked electron
spectra from dielectric solids, may thus be applied to probe solid
state characteristic, in particular, the solid's ultra fast
dielectric response to a moving charge with unprecedented accuracy.
This is supported by our quantitative prediction of wake-induced
delays exceeding 50~as that fall within the temporal resolution
achievable with contemporary laser technology~\cite{Schultze10}.

This work was supported by the NSF, the Division of Chemical
Sciences, Office of Basic Energy Sciences, Office of Energy
Research, US~DOE, and access to the Beocat computer cluster
at Kansas State University.

\bibliographystyle{apsrev}
\bibliography{attosecond_con}

\begin{thebibliography}{27}
\expandafter\ifx\csname natexlab\endcsname\relax\def\natexlab#1{#1}\fi
\expandafter\ifx\csname bibnamefont\endcsname\relax
  \def\bibnamefont#1{#1}\fi
\expandafter\ifx\csname bibfnamefont\endcsname\relax
  \def\bibfnamefont#1{#1}\fi
\expandafter\ifx\csname citenamefont\endcsname\relax
  \def\citenamefont#1{#1}\fi
\expandafter\ifx\csname url\endcsname\relax
  \def\url#1{\texttt{#1}}\fi
\expandafter\ifx\csname urlprefix\endcsname\relax\def\urlprefix{URL }\fi
\providecommand{\bibinfo}[2]{#2}
\providecommand{\eprint}[2][]{\url{#2}}

\bibitem[{\citenamefont{Harris and Jones}(1973)}]{Harris73}
\bibinfo{author}{\bibfnamefont{J.}~\bibnamefont{Harris}} \bibnamefont{and}
  \bibinfo{author}{\bibfnamefont{R.~O.} \bibnamefont{Jones}},
  \bibinfo{journal}{J. Phys. C} \textbf{\bibinfo{volume}{6}},
  \bibinfo{pages}{3585} (\bibinfo{year}{1973}).

\bibitem[{\citenamefont{Manson and Ritchie}(1981)}]{Manson81}
\bibinfo{author}{\bibfnamefont{J.~R.} \bibnamefont{Manson}} \bibnamefont{and}
  \bibinfo{author}{\bibfnamefont{R.~H.} \bibnamefont{Ritchie}},
  \bibinfo{journal}{Phys. Rev. B} \textbf{\bibinfo{volume}{24}},
  \bibinfo{pages}{4867} (\bibinfo{year}{1981}).

\bibitem[{\citenamefont{Echenique et~al.}(1990)\citenamefont{Echenique, Flore,
  and Ritchie}}]{Echenique90}
\bibinfo{author}{\bibfnamefont{P.~M.} \bibnamefont{Echenique}},
  \bibinfo{author}{\bibfnamefont{F.}~\bibnamefont{Flore}}, \bibnamefont{and}
  \bibinfo{author}{\bibfnamefont{R.~H.} \bibnamefont{Ritchie}},
  \bibinfo{journal}{Solid State Physics} \textbf{\bibinfo{volume}{43}},
  \bibinfo{pages}{229} (\bibinfo{year}{1990}).

\bibitem[{Gar()}]{Garcia92}
\bibinfo{note}{F. J. Garc\'{\i}a de Abajo and P. M. Echenique, Phys. Rev. B
  {\bf 46}, 2663(1992); Phys. Rev. B {\bf 48}, 13399(1993)}.

\bibitem[{Luc()}]{Lucas70}
\bibinfo{note}{A. A. Lucas, E. Karthueser, and R. C. Badro, Phys. Rev. B {\bf
  2}, 2488 (1970)}.

\bibitem[{Sun()}]{Sunjic71}
\bibinfo{note}{M. \^{S}unji\'{c} and A. A. Lucas, Phys. Rev. B {\bf 3}, 719
  (1971)}.

\bibitem[{Ove()}]{Overhauser71}
\bibinfo{note}{A. W. Overhauser, Phys. Rev. B {\bf 3}, 1888 (1971)}.

\bibitem[{Bur()}]{Burgdorfer}
\bibinfo{note}{J. Burgd\"orfer, Nucl. Instrum. Methods {\bf 24/25}, 139 (1987);
  C. O. Reinhold and J. Burgd\"orfer, Phys. Rev. A {\bf 55}, 450 (1997)}.

\bibitem[{Ger()}]{GersTzoar73}
\bibinfo{note}{J. I. Gersten and N. Tzoar, Phys. Rev. B {\bf 8}, 5671 (1973),
  N. Tzoar and J. I. Gersten, Phys. Rev. B {\bf 8}, 5684 (1973)}.

\bibitem[{Sch()}]{Schultze10}
\bibinfo{note}{M. Schultze {\em et al.}, Science {\bf 328}, 1658 (2010)}.

\bibitem[{Cav()}]{Cavalieri07}
\bibinfo{note}{A. L. Cavalieri {\em et al.}, Nature {\bf 449}, 1029 (2007)}.

\bibitem[{Dre()}]{Drescher02}
\bibinfo{note}{M. Drescher {\em et al.}, Nature {\bf 419}, 803 (2002)}.

\bibitem[{Mia()}]{Miaja08}
\bibinfo{note}{L. Miaja-Avila {\em et al.}, Phys. Rev. Lett. {\bf 101}, 046101
  (2008)}.

\bibitem[{Kub()}]{Kubo05}
\bibinfo{note}{A. Kubo {\em et al.}, Nano Lett. {\bf 5}, 1123 (2005)}.

\bibitem[{Sto()}]{Stockman07}
\bibinfo{note}{M.I. Stockman {\em et al.}, Nature Phot. {\bf 1}, 539 (2007)}.

\bibitem[{Suk()}]{Sukharev06}
\bibinfo{note}{M. Sukharev and T. Seideman, Nano Lett. {\bf 6}, 1123 (2006)}.

\bibitem[{Nor()}]{Nordlander07}
\bibinfo{note}{F. Le {\em et al.}, Phys. Rev. B {\bf 76}, 165410 (2007); F. Hao
  {\em et al.}, Phys. Rev. B {\bf 76}, 165410 (2007)}.

\bibitem[{Zha({\natexlab{a}})}]{Zhang09}
\bibinfo{note}{C.-H. Zhang and U. Thumm, Phys. Rev. Lett. {\bf 102}, 123601
  (2009)}.

\bibitem[{Zha({\natexlab{b}})}]{Zhang10}
\bibinfo{note}{C.-H. Zhang and U. Thumm, Phys. Rev. A {\bf 82}, 043405 (2010)}.

\bibitem[{Bag()}]{Baggesen09}
\bibinfo{note}{J. C. Baggesen and L. B. Madsen, Phys. Rev. A {\bf 78} 032903
  (2008); ibid. {\bf 80} 030901(R) (2009)}.

\bibitem[{Lem()}]{Lemell09}
\bibinfo{note}{C. Lemell, B. Solleder, K. T\"{o}k\'{e}si, and J.
  Burgd\"{o}rfer, Phys. Rev. A {\bf 79}, 062901 (2009)}.

\bibitem[{Kaz()}]{Kazansky09}
\bibinfo{note}{A. K. Kazansky and P. M. Echenique, Phys. Rev. Lett. {\bf 102}
  177401 (2009)}.

\bibitem[{\citenamefont{Echenique et~al.}(1981)\citenamefont{Echenique,
  Ritchie, Barber\`{a}n, and Inkson}}]{Echenique81}
\bibinfo{author}{\bibfnamefont{P.~M.} \bibnamefont{Echenique}},
  \bibinfo{author}{\bibfnamefont{R.~H.} \bibnamefont{Ritchie}},
  \bibinfo{author}{\bibfnamefont{N.}~\bibnamefont{Barber\`{a}n}},
  \bibnamefont{and} \bibinfo{author}{\bibfnamefont{J.}~\bibnamefont{Inkson}},
  \bibinfo{journal}{Phys. Rev. B} \textbf{\bibinfo{volume}{23}},
  \bibinfo{pages}{6486} (\bibinfo{year}{1981}).

\bibitem[{\citenamefont{Ibach}(2006)}]{Ibach06}
\bibinfo{author}{\bibfnamefont{H.}~\bibnamefont{Ibach}},
  \emph{\bibinfo{title}{Physics of Surfaces and Interfaces}}
  (\bibinfo{publisher}{Springer-Verlag}, \bibinfo{address}{Berlin},
  \bibinfo{year}{2006}).

\bibitem[{\citenamefont{Ashcroft and Mermin}(1976)}]{Ashcroft76}
\bibinfo{author}{\bibfnamefont{N.~W.} \bibnamefont{Ashcroft}} \bibnamefont{and}
  \bibinfo{author}{\bibfnamefont{N.~D.} \bibnamefont{Mermin}},
  \emph{\bibinfo{title}{Solid State Physics}} (\bibinfo{publisher}{Thomson
  Learning}, \bibinfo{address}{USA}, \bibinfo{year}{1976}).

\bibitem[{not()}]{notez}
\bibinfo{note}{The temporal shift calculated with $\delta U^i$ is qualitatively
  equal to that obtained with the phenomenological imaginary potential $\delta
  U^i_{ph}$ for $\lambda=5$~\AA. Details on the quality of calculations with
  $\delta U^i_{ph}$ will be published separately.}

\bibitem[{Car()}]{Carvalho02}
\bibinfo{note}{C. A. A. de Carvalho and H. M. Nussenzweig, Phys. Rep. {\bf 364}
  83 (2002)}.

\end{thebibliography}

\end{document}